\pdfoutput=1
\documentclass[prx,aps,twocolumn,superscriptaddress,longbibliography]{revtex4-1}

\usepackage{graphicx}
\usepackage{dcolumn}
\usepackage{bm}
\usepackage{multirow}
\usepackage{gensymb}

\newcommand{\Tone}{\ensuremath{T_1}} 
\newcommand{\Ttwo}{\ensuremath{T_2}} 
\newcommand{\um}{\ensuremath{\mu\mathrm{m}}} 
\newcommand{\umsq}{\ensuremath{\mu\mathrm{m}^2}} 
\newcommand{\us}{\ensuremath{\mu\mathrm{s}}} 
\newcommand{\ejec}{\ensuremath{E_{J}/E_{C}}}
\newcommand\T{\rule{0pt}{2.6ex}}       
\newcommand\B{\rule[-1.2ex]{0pt}{0pt}} 
\newcommand{\AlOx}{AlO$_{x}$}
\newcommand{\kohm}{k$\Omega$}

\begin{document}

\title{Merged-Element Transmons: Design and Qubit Performance}


\author{H. J. Mamin}
\email{mamin@us.ibm.com}
\author{E. Huang}
\email{ehuang@us.ibm.com}
\affiliation{IBM Quantum, IBM Almaden Research Center, San Jose, CA 95120, USA}
\author{S. Carnevale}
\affiliation{IBM Quantum, IBM T. J. Watson Research Center, Yorktown Heights, NY 10598, USA}
\author{C. T. Rettner}
\author{N. Arellano}
\author{M. H. Sherwood}
\affiliation{IBM Quantum, IBM Almaden Research Center, San Jose, CA 95120, USA}
\author{C. Kurter}
\author{B. Trimm}
\author{M. Sandberg}
\affiliation{IBM Quantum, IBM T. J. Watson Research Center, Yorktown Heights, NY 10598, USA}
\author{R. M. Shelby}
\author{M. A. Mueed}
\author{B. A. Madon}
\author{A. Pushp}
\affiliation{IBM Quantum, IBM Almaden Research Center, San Jose, CA 95120, USA}
\author{M. Steffen}
\affiliation{IBM Quantum, IBM T. J. Watson Research Center, Yorktown Heights, NY 10598, USA}
\author{D. Rugar}
\email{rugar@us.ibm.com}
\affiliation{IBM Quantum, IBM Almaden Research Center, San Jose, CA 95120, USA}

\date{\today}

\begin{abstract}

We have demonstrated a superconducting transmon qubit in which a Josephson junction has been engineered to act as its own parallel shunt capacitor. This merged-element transmon (MET) potentially offers a smaller footprint than conventional transmons. Because it concentrates the electromagnetic energy inside the junction, it reduces relative electric field participation from other interfaces. By combining micrometer-scale Al/AlO$_{x}$/Al junctions with long oxidations, we have produced functional devices with \ejec\ in the low transmon regime ($\ejec \lesssim 30$). Cryogenic I-V measurements show sharp $dI/dV$ structure with low sub-gap conduction. Qubit spectroscopy of tunable versions show a small number of avoided level crossings, suggesting the presence of two-level systems (TLS). We have observed mean \Tone\  times typically in the range of 10-90~\us, with some annealed devices exhibiting $\Tone > 100$~\us\ over several hours. The results suggest that energy relaxation in conventional, small-junction transmons is not limited by junction loss.
\end{abstract}

\maketitle


\section{\label{sec:intro}Introduction}

The superconducting transmon qubit \cite{koch_transmon} has become a workhorse in the field of quantum computation and is the fundamental building block in some of the most sophisticated quantum computation systems built to date \cite{jurcevic2020demonstration,aleiner2020accurately}. The transmon consists of a Josephson junction (JJ) in parallel with a coplanar shunt capacitor, forming a simple nonlinear LC circuit. The shunt capacitor acts to exponentially suppress charge noise while retaining enough anharmonicity to allow individual quantized transitions to be addressed. In principle, it could be made simpler by engineering the junction self-capacitance to be large enough to act as its own shunt capacitor, eliminating the need for an external capacitor \cite{Tahan,zhao2020mergedelement}. Such qubits could also be significantly more compact, allowing for higher areal density of qubits. Moreover, because they concentrate the energy inside the junction, the relative importance of other lossy interfaces and surfaces should be reduced. This could conceivably lead to improved coherence if high quality (for example, epitaxial) dielectrics can be developed \cite{Tahan,PappasEpitaxy,nakamura2011epitaxial}. This concept has been dubbed the Merged-Element Transmon, or MET \cite{zhao2020mergedelement}. 

\section{\label{sec:design}Design and Simulation}

The tunnel junction at the heart of the MET does double duty as a Josephson element and a parallel plate capacitor. The dimensions are constrained by the target capacitance as well as the thickness and dielectric constant of the insulating tunnel barrier. The exact dielectric thickness is not known \textit{a~priori}, but must be in a limited range, given the exponential dependence of critical current on thickness. Accordingly, as a starting point we assumed an oxide thickness of 2~nm with a dielectric constant of 10, appropriate for the Al/AlO$_{x}$/Al tunnel junctions that we used for these initial studies. Applying  the formula for a simple parallel plate capacitor gives a junction area of roughly $1.4~\um^2$ to achieve a junction capacitance of 62~fF. In principle, since the junction area and the qubit area could be one and the same, the qubit footprint could be greatly reduced compared to conventional transmons that use coplanar capacitors with dimensions of hundreds of micrometers \cite{Gambetta_IEEE}. In terms of junction dimensions, the MET falls somewhere between a transmon, with sub-micron junction dimensions, and phase qubits, which are  self-shunting like the MET, but are typically much larger laterally and capacitively, increasing the probability of being plagued by two level systems (TLS)\cite{Phase_qubit_Rabi,PhysRevLett.95.210503}. 

The MET geometry results in two significant challenges related to the fact that the MET lateral dimensions are large for a junction but small for a capacitor. On the one hand, the small capacitor geometry gives very little area for capacitive coupling to the drive/readout circuitry. On the other hand, the junction area is up to two orders of magnitude larger than typical transmon junctions. To achieve the same critical current needed for 4-5 GHz qubit operation (typically $\sim 24$~nA) therefore requires making the tunnel barrier appropriately thicker.

\begin{figure}
\begin{center}
\includegraphics[width=0.45\textwidth]{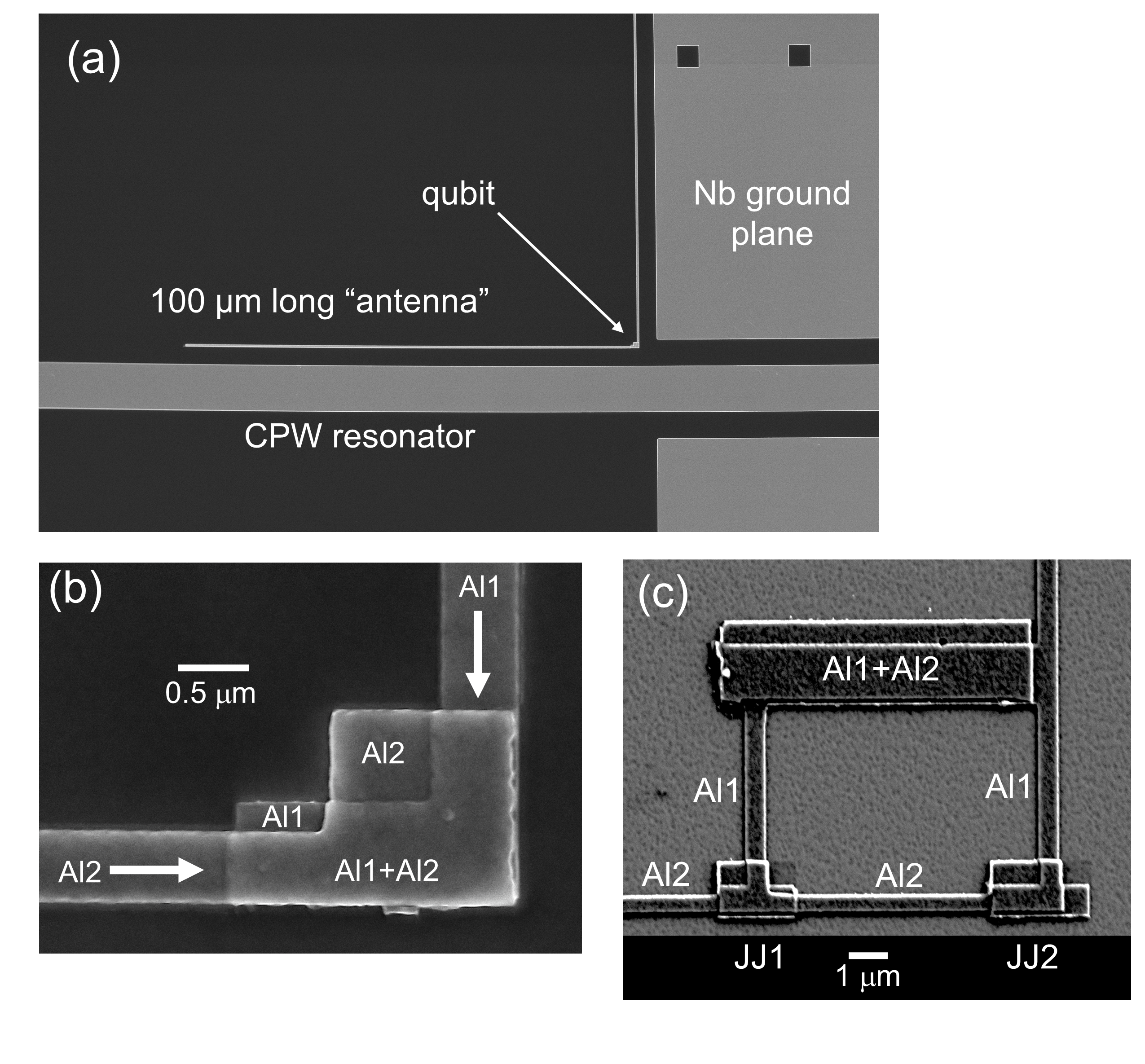}
\end{center}
\caption{Scanning electron micrographs of MET devices. (a) The MET consists of a micrometer-size tunnel junction capacitively coupled to a coplanar waveguide resonator via 100~\um -long, 0.5~\um -wide ``antenna" structures. The resonator, which serves to both excite the qubit and read out its state, has a resonant frequency around 7~GHz. (b) Close-up of the junction region. The various levels and shapes are a consequence of the angled evaporations and shadowing by the resist layer. Azimuthal directions of the two evaporations are indicated by arrows. The junction area is determined by the overlap of Al1 and Al2, which in this case is approximately 1.4~\umsq. (c) A two-junction device that serves as a flux tunable qubit.}   
\label{fig:Fig1abc}
\end{figure}

Assuming that a junction of proper capacitance could be made, the chief design decision became how to couple to it. For the present work, we chose to use coplanar capacitive coupling to the readout/drive resonator for simplicity, with a target coupling strength $g/2\pi$ of at least 20 MHz to obtain an adequate readout signal-to-noise ratio. An additional constraint was that the coplanar coupling structure should not add substantially to the overall capacitance so as not to detract unduly from the self-shunting nature of the junction. Using finite element simulations (ANSYS Q3D) to guide the design, we satisfied the constraints using the geometry shown in Figs.~\ref{fig:Fig1abc}(a) and (b). Two 100~\um \ long arms with width of 0.5~\um \ extend out perpendicularly from the junction region; these arms act as antennas and allow for coupling to a readout resonator and ground plane respectively. The coplanar contribution from just the arms to the overall capacitance, calculated by removing the overlapping junction region, was 5.0 fF. Thus, with a junction capacitance of $\sim$62~fF, the junction was responsible for 93\% of the overall device capacitance (Table~\ref{tab:table0}). In future designs, the 7\% contribution from the arms could be made even less by extending the arms out at a 180\degree\ instead of 90\degree\ angle.  While our geometry gives up a substantial amount of the MET’s areal density advantage, with the junction itself representing only a small fraction of the device footprint, it is still considerably smaller than many other transmon designs \cite{Gambetta_IEEE, Corcoles2015}.

This approach successfully solved the coupling problem, with coupling factor $g/2\pi$ predicted to be roughly 27 MHz for the antennas spaced 4~\um \ from the readout resonator and ground plane. The coupling could in principle be made more compact by decreasing the spacing of the antenna arms from the resonator and ground, or by creating interdigitated structures. 
However, finite element simulations showed that significant substrate-vacuum participation was associated with the coupling regions, where high electric fields exist. Compared to a typical conventional transmon with a coplanar shunt capacitor, such as the ``mod D" design in \cite{Gambetta_IEEE}, the substrate-vacuum participation was calculated to be reduced only by roughly 30 percent, setting an upper limit on the coherence gains that might be observed if the substrate-vacuum interface is particularly lossy. Going to a parallel plate-type coupler \cite{zhao2020mergedelement}, though not as simple to fabricate, might allow for higher coupling with reduced surface participation and footprint. 

\begin{table*}[tb]
\caption{Comparison of typical qubit properties. The MET entries are initial targeted values. Properties for conventional transmons represent calculated and inferred values for typical geometries as given in the references.}
    \label{tab:table0}
  \centering

\begin{center}
 \begin{tabular}{|c|c|c|}
 \hline
 \hline
Property &\ \ \  MET \ \ \  & \ \  Conventional Transmon \cite{Gambetta_IEEE,hertzberg2020laser,Place2021} \ \ \T \B \\
 \hline
Junction area (\umsq) & 1.4 & 0.01-0.03  \T \\
Junction capacitance, $C_{\mathrm{JJ}}$ (fF) & 62 & 0.5-1.3 \\
Total capacitance, $C_{\mathrm{total}}$ (fF) & 67 & 60 \\
Junction participation, $p_{\mathrm{JJ}}$ & 0.93 & 0.008-0.02 \\
Substrate-vacuum participation {(nm$^{-1}$)  } &\  $3.5 \times 10^{-5}$ \ & $5.0 \times 10^{-5}$  \B \\
 \hline
 \hline
\end{tabular}

\end{center}

\end{table*}

\section{\label{sec:fab}Fabrication}

For MET fabrication, we aimed for a total capacitance of 67~fF and critical current of 24~nA to produce qubits with frequency around 5~GHz, anharmonicity of 350~MHz and  \ejec \ ratio near 40 \cite{koch_transmon}. As the optimal junction area and oxidation conditions were not known initially, devices with various junction areas and oxidations were fabricated to explore electrical characteristics and qubit performance. 
 
All devices were fabricated on intrinsic, high resistivity (100) silicon wafers. Prior to junction fabrication, niobium structures (e.g., coplanar waveguide resonators) were fabricated using a standard optical lithography process ($\lambda$ = 248~nm), followed by reactive ion etching. 


Junctions with areas ranging from 1 to 2.4 \umsq \ were fabricated using a variation of a bridgeless ``Manhattan" approach \cite{ManhattanUCB,Manhattan_Costache,Manhattan_Potts,Zhang}.
In this approach, an electron-beam lithography pattern was defined in a 660~nm thick positive-tone bilayer resist with narrow regions defining the antenna arms and a larger overlap region where the junction is to be formed. An initial 50~nm thick aluminum deposition (Al1) was performed at a 45 degree incident polar angle, with the azimuthal direction aligned along the direction of one of the antenna arms having a width of 500~nm (Fig.~\ref{fig:Fig1abc}(b)). This Al1 deposition thus forms one arm as well as the base electrode of the junction. No aluminum is deposited along the perpendicular arm due to the narrowness of the antenna pattern and the shadowing effect of the resist stack.

The sample was then moved to a separate chamber without breaking vacuum for oxidation in order to form the tunnel barrier. To achieve the desired junction critical current, a relatively thick tunnel barrier was needed, which required rather long oxidations at relatively high oxygen pressure. Typical values were 1-4 hours at 600~Torr of O$_{2}$ (more details below). 

After oxidation, the wafer was rotated by an azimuthal angle of 90 degrees for the second deposition. This deposition of aluminum (Al2) was 100 nm thick and formed both the counter electrode and the second antenna arm. Subsequently, the devices were exposed to a solvent strip to remove the bilayer resist and lift off the Al1/Al2 layers residing atop the resist. The resulting structure is shown in Fig.~\ref{fig:Fig1abc}(b), where the overlap region of the two depositions formed the tunnel junction.

In addition to single junction devices, we also fabricated a two junction version of the MET in order to have a flux tunable device. The finished device, shown in Fig.~\ref{fig:Fig1abc}(c), has two equal-area Josephson junctions connected in parallel to form a SQUID loop configuration. Due to the fabrication process, this two junction device has a parasitic junction, seen as the larger rectangle at the top of the micrograph where there is an overlap of the base and counter electrode layers. 

Since MET performance is dominated by the quality of the tunnel junction, a number of oxidation conditions and heat treatments were considered in order to optimize and tune the tunnel junction characteristics. Based on experience with conventional transmons, we initially sought room temperature junction resistances in the neighborhood of 10~\kohm\ in order to achieve Josephson critical currents around 24~nA. Using devices with 2.4 \umsq \ junction area, an initial test with a 1~hour oxidation at 600~torr resulted in resistance values that were below our target, in the vicinity of 5.3~\kohm. Previous studies have shown that heat treatments in the range of 350 - 450\degree C can increase the junction resistance as well as improve junction quality \cite{Scherer_JJanneal_2001, Koppinen_JJanneal_2007, Julin_JJanneal_2010, pop2012fabrication}. Accordingly, we tested the effect of rapid thermal anneal in a nitrogen atmosphere. Anneals for 5~minutes at both 375 and 425\degree C were found to increase the room temperature resistance to approximately 6.2~\kohm. 

\begin{figure}
\begin{center}
\includegraphics[width=0.37\textwidth]{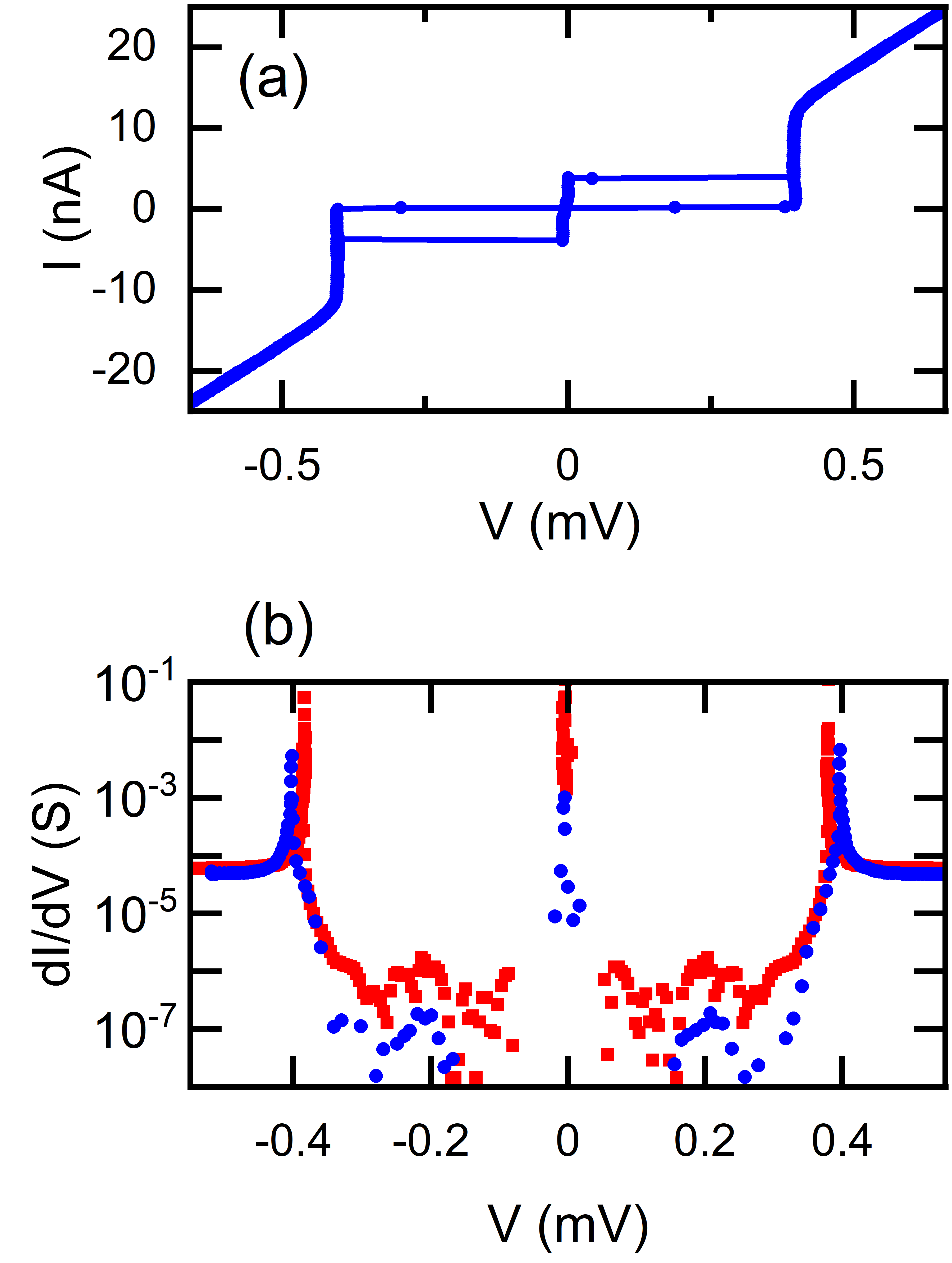}
\end{center}
\caption{Electrical characteristics of tunnel junctions with 1~\umsq \ junction area taken at 20~mK. Junction oxidation was 1~hour at 600~torr. (a) DC current-voltage characteristics. (b) $dI/dV$ measurement of junctions showing low sub-band gap conductivity. The noisy data at the lowest conductance levels are limited by instrumental resolution. The blue data points are for a device annealed at 425\degree C for five minutes. It exhibits a slightly larger energy gap compared to the unannealed device (red data).}   
\label{fig:MET_IV}
\end{figure}

To confirm that the fabrication process produced high quality tunnel junctions, DC current-voltage characteristics were measured at millikelvin temperature \cite{IVsetup}. Both annealed and unannealed devices exhibited low sub-gap conduction and sharp turn-on at the superconducting gap, as demonstrated by the $I$-$V$ and $dI/dV$ curves shown in Fig.~\ref{fig:MET_IV}. For devices subjected to the 425\degree C anneal (blue curve in Fig.~\ref{fig:MET_IV}(b)), a small increase in the superconducting gap was evident compared to the unannealed device.  Gap values of 200~$\mu$eV and 191~$\mu$eV were found for the annealed and unannealed devices, respectively, as determined by fitting the peaks in the differential conductance to the BCS model \cite{tinkham2004introduction}. Somewhat lower sub-gap conductivity was also observed for the annealed junction, indicative of improved junction quality \cite{gubrud2001sub}. 

For the qubit results presented below, the oxidation time was increased to 4~hours at 600~torr in order to further increase the tunneling resistance. The resulting room temperature resistances were approximately 6.6~\kohm\ for unannealed junctions and 9.0~\kohm\ for junctions annealed at 425\degree C. For the unannealed devices a brief argon ion milling step was performed just prior to the deposition of Al1. This was found to improve the yield of the unannealed qubits. 

\begin{table*}[tb]
    \caption{Characteristics of selected qubits.}
    \label{tab:table1}
  \centering

\begin{center}
 \begin{tabular}{|c|c|c|c|c|ccc|ccc|c|c|}
 \hline
 \hline
 Qubit & JJ Area & $f_{01}$ & $\alpha/2\pi$  &  \multirow{2}*{ $E_j/E_c$  } & \multicolumn{3}{c|}{\ $T_1$ ($\mu$s) \ } & \multicolumn{3}{c|}{$T_{2}$-echo ($\mu$s)} & \  Mean $Q$ \  &\multirow{2}*{ Type } \T \\
ID & (\umsq) & \ (GHz) \  & \  (MHz) \  & &  \ Best \  & \  Mean \ \ & Std. dev. & \ Best \  & \ Mean \ \ & \ Std.~dev. & (M)  & \B \\
 \hline
 \rule{0pt}{3ex}J4 & 1.4 & 3.808 & 414 & 21 & 234 & 89.9 & 75.9 & - & - & -  & 2.2 & \multirow{5}*{Annealed} \\
 K7 & 1.9 & 3.747 & 343 & 27 & 109 & 88.1 & 12.6 & 50 & 41.1 & 3.8  & 2.1 & \\
 J7 & 1.9 & 3.748 & 362 & 25 & 154 & 87.4 & 52.5 & - & - & -  & 2.1 &  \\
 K5 & 1.9 & 3.771 & 339 & 27 & 65 & 50.3 & 8.7 & 39 & 33.4 & 3.1  & 1.2 & \\
 J6 & 1.9 & 3.758 & 368 & 24 & 41 & 38.1 & 1.6 & 46 & 43.3 & 2.2  & 0.90 &  \\
 \hline
 \rule{0pt}{3ex}A6 & 1.9 & 4.978 & 404 & 32 & 41 & 34.4 & 3.9 & 28 & 21.1 & 2.0 & 1.1 & \multirow{5}*{Unannealed} \\
 B9 & 1.4 & 4.521 & 439 & 25 & 23 & 16.9 & 6.4 & - & - & - & 0.48 &  \\
 A9 & 1.4 & 4.610 & 426 & 26 & 29 & 16.1 & 8.6 & - & - & - & 0.47 &  \\
 A5 & 1.9 & 5.032 & 417 & 31 & 32 & 14.6 & 6.2 & 32 & 20.5 & 6.8 & 0.46 &  \\
 B7 & 1.9 & 4.503 & 376 & 30 & 15 & 11.8 & 2.3 & - & - & - & 0.33 &  \\
 \hline
 \hline
\end{tabular}
\end{center}
\end{table*}

\section{\label{sec:characterization}Qubit Characterization}

Single-junction MET qubits with junction areas of 1.4 and 1.9 \umsq \  were characterized in a well-shielded dilution refrigerator operating below 20~mK \cite{supplement}. The qubits were capacitively coupled to quarter-wave coplanar waveguide resonators, which were, in turn, inductively coupled to a 50~ohm feedline for transmission-mode dispersive readout. Typical resonator frequencies were around 7~GHz. Functionality was evaluated using both continuous wave (cw) and pulsed microwave excitation for unannealed and annealed devices. Table~\ref{tab:table1} summarizes the results for the five best performing devices of each type. 


 Two-tone cw spectroscopy measurements using a vector network analyzer were used to determine the $f_{01}$ qubit frequency as well as the anharmonicity $\alpha/2\pi=f_{01}-f_{12}=2f_{01}-f_{02}$ \cite{zhao2020mergedelement}. Qubit frequencies were typically in the range 4.4 - 5.0~GHz for unannealed devices and 3.3 - 3.8 GHz for the annealed devices. The lower frequencies for the annealed devices are due to the lower critical current and larger Josephson inductance of the heat treated junctions. Anharmonicities typically ranged from 300 - 450~MHz. From these measurements we can calculate \ejec, the ratio of Josephson energy to charging energy \cite{koch_transmon}. The devices with the larger junction area (1.9~\umsq) had somewhat smaller anharmonicities and larger \ejec \ ratios, as expected from their higher capacitance. Overall, the \ejec \ ratios were mostly in the range of 20-30, which was somewhat lower than targeted. Presumably this low ratio was responsible for significant charge noise observed in some of the qubits.
 
\begin{figure}[t]
\begin{center}
\includegraphics[width=0.45\textwidth]{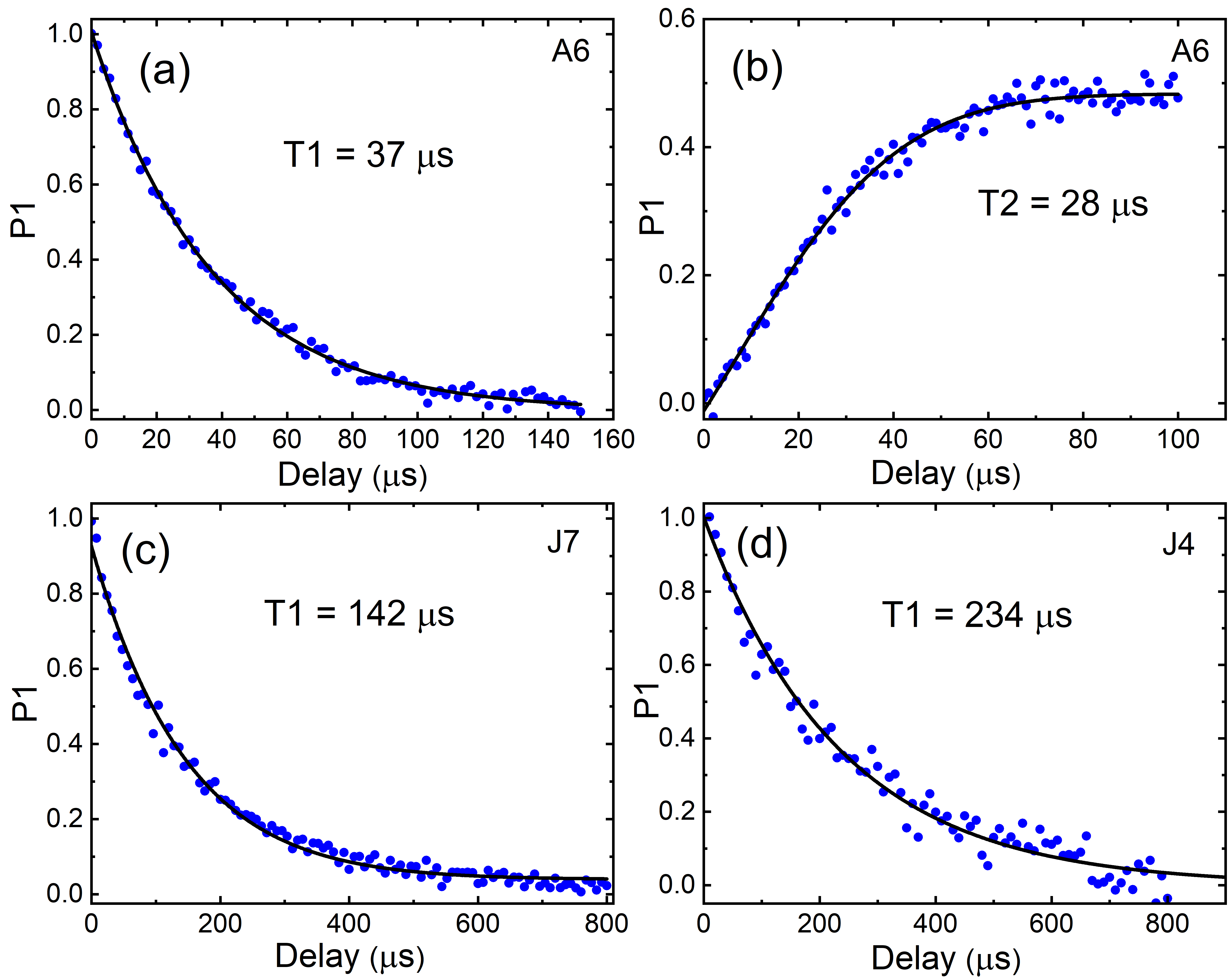}
\end{center}
\caption{Examples of measured \Tone \ and $T_2$-echo decay curves plotted as excited state probability $P1$ vs. readout delay time. Curves (a) and (b) are for unannealed qubit A6. Curves (c) and (d) are for annealed qubits J7 and J4, respectively. Solid lines are exponential fits, except for (b), where a stretched exponential of the form $A+B\exp[-(t/T_2)^n]$ was used, with n = 1.37. }  
\label{fig:T1T2}
\end{figure}

Using pulsed time-domain sequences, we successfully measured both the energy relaxation time \Tone\  and the echo decoherence time \Ttwo\ for a number of qubits \footnote{The main emphasis of our study was the \Tone\ performance. Not all qubits were measured for \Ttwo -echo}.  Figure~\ref{fig:T1T2} and Table~\ref{tab:table1} show results obtained for some of the better performing devices. The best performing unannealed device had a mean \Tone\ of 34.4~\us\  when averaged over several hours with 87 separate measurements. Overall, the median \Tone\ for the 14 unannealed devices we measured was 13~\us, with a median qubit quality factor $Q = 2 \pi f_{01} \Tone$ of $3.8\times 10^5$.  

The annealed devices performed considerably better. The median \Tone\ for the eight annealed devices we measured was 46~\us, with a median $Q$ of $1.1\times 10^6$. Three of the best performing annealed devices had mean \Tone\ greater than 87~\us, corresponding to quality factors above 2 million. Remarkably, one annealed qubit maintained a mean \Tone\  greater than 200~\us\  (Fig.~\ref{fig:T1T2}(d)) over a period of several hours before abruptly dropping down to more typical values. Similar \Tone\ fluctuations have been seen previously in conventional transmon qubits and are believed to be due to two-level systems coming into resonance with the qubit \cite{Klimov_fluctuations}.

\section{\label{sec:spectroscopy}Qubit spectroscopy}

\begin{figure}[b]
\begin{center}
\includegraphics[width=0.45\textwidth]{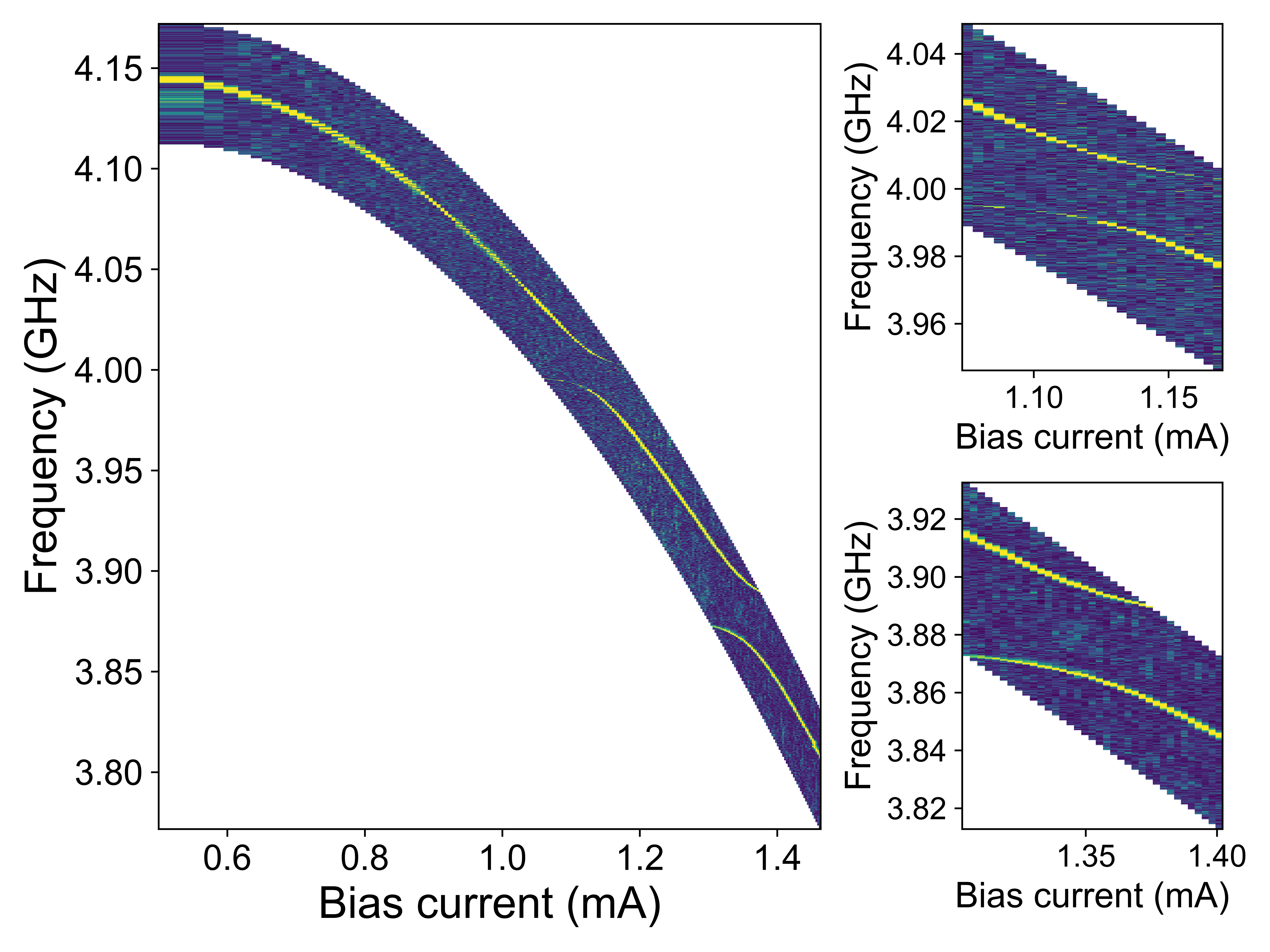}
\end{center}
\caption{Qubit spectroscopy showing qubit frequency as a function of magnet bias current in a flux tunable MET device. Detail on the right shows two prominent avoided crossings  with frequency splittings on the order of 20 MHz, suggesting the presence of two-level systems within the MET junction. }   
\label{fig:colormap}
\end{figure}

Because roughly 90 percent of the electromagnetic energy is confined to the junction, the MET is an ideal testbed for studying the properties of the dielectric layer and losses in the junction. One approach is to perform qubit spectroscopy to look for signs of individual two-level systems (TLS). This is most readily done with our flux tunable, two-junction MET (Fig.~\ref{fig:Fig1abc}(c)). A small coil electromagnet was attached to the qubit board to tune the qubit frequency. This allowed for two-dimensional qubit spectroscopy, where both the coil current (magnetic flux) and qubit pump frequency were varied. 
The resulting false color plot for an annealed device is shown in Fig.~\ref{fig:colormap}.  Two prominent avoided crossings are seen with splittings of 20 to 30 MHz, similar to what has been seen in phase qubits \cite{PhysRevLett.95.210503,MartinisAvoidedCrossingsPhysRevLett.93.180401,Dielectric_loss}. 

We observed avoided crossings in both annealed and unannealed devices. In general, the avoided crossings were rare over the measured $\sim$1~GHz frequency range, but, because the scans were rather coarse, it is possible some number of smaller splittings went undetected. Nonetheless, we can naively use these measurements to get an order-of-magnitude estimate of the density of strongly coupled TLS in our junctions (e.g., coupling strength $>$10~MHz). Combining results from both annealed and unannealed devices, we detected a total of 17 avoided crossings over six qubits, giving an average density of $1.0 \  \um^{-2}\,\mathrm{GHz}^{-1}$ \footnote{Each of the six tunable devices had a total junction area of $2.9 \ \um^{2}$ and an average measured frequency range of 1~GHz. The possible effect of the parasitic junction in the tunable devices was not taken into account.}. 
This value is in reasonable agreement with the value of $\sim 0.5 \ \um^{-2}\,\mathrm{GHz}^{-1}$ from Martinis \textit{et~al.}, derived from measurements on larger junctions, as well as values from other bulk and thin film dielectrics \cite{PhysRevLett.95.210503,Lisenfeld2019}. Further studies will be needed to determine if there is a statistically significant difference between annealed and unannealed devices. Ultimately, reducing the TLS density will require more perfect tunnel barriers, such as made through epitaxial means. 

\section{\label{sec:implications}Implications for Conventional Small-Junction Transmons}

Given that the performance of the MET will be dominated by the loss in the junction, what does the MET tell us about junction loss in general, and can this knowledge be used to make inferences about conventional small-junction transmons? Using scaling arguments and some simplifying assumptions, one can in fact argue that the junction must not be the dominant source of energy loss in conventional small-junction transmons. 

We start by writing the total loss $\Gamma_{Q}$ as the sum of individual loss terms \cite{Wang2015}:
\begin{equation}
\Gamma_{\mathrm {Q}} = p_{\mathrm{JJ}} \tan \delta_{\mathrm{JJ}} + \sum_i p_i \tan \delta_i 
\end{equation}
where $p_{\mathrm{JJ}}$ is the fraction of the electric field energy associated with the Josephson junction, $p_i$ represents the fraction of energy in various other materials and interfaces, and $\tan \delta_i$ is the associated loss tangent. 

The energy in the junction is just ${1 \over 2} C_{\mathrm{JJ}} V^2$, while the total energy is ${1 \over 2} C_{\mathrm{total}} V^2$. Thus $p_{\mathrm{JJ}} = C_{\mathrm{JJ}} / C_{\mathrm{total}}$. For the MET, virtually all the capacitance is due to the junction, i.e. $p_{\mathrm{JJ}} \sim 0.93$. A conventional, small-junction transmon will have much smaller capacitance and junction participation based on its smaller area. A transmon with a junction area of 0.03~$\um^2$, for example \cite{Place2021}, would be expected to have $p_{\mathrm{JJ}}$ about 46 times smaller than for the 1.4~\umsq \ MET. Here we make the simplifying assumption that both types of junction have roughly the same oxide thickness and loss tangent. (We note, however, that this assumption may not be valid, since the oxidation conditions are different.) Equation (1) therefore implies that the junction loss term for this device should be 46 times smaller than for the MET. Based on the best mean \Tone \ from Table II of 89.9~\us, the limit on \Tone \ imposed by the junction in a conventional transmon would then be 4.1~ms, assuming only junction loss. To the extent that such long relaxation times are not observed in conventional transmons, sources of loss other than the junction must be limiting the coherence. This result is consistent with the conclusion reached by Wang \textit{et al.} \cite{Wang2015} based on a study of qubit relaxation as a function of electric field surface participation. 

Alternatively, we can use our MET results to determine an upper bound to the loss tangent of the \AlOx \ in the junction.  We see from Eq. (1) that 
\begin{equation}
 \tan \delta_{\mathrm{JJ}} < \Gamma_{\mathrm {Q}}/p_{\mathrm {JJ}}.
\end{equation}
Since $\Gamma_{Q}=1/Q$, where $Q$ is the qubit quality factor, and $p_{\mathrm {JJ}}\sim 0.93$, this implies that $\delta_{\mathrm{JJ}} \lesssim 1/0.93\, Q$. With a best measured mean $Q$ of $2.2 \times 10^6$, we find $\delta_{\mathrm{JJ}}\lesssim 5 \times 10^{-7}$.

While this loss tangent is small compared to typical literature values for \AlOx, where loss tangents on the order of $10^{-3}$ are commonly found \cite{PhysRevLett.95.210503, Pappas_IEEE_TLS_loss_2011, Deng_AlOx_loss_APL_2014}, larger values would not be compatible with the measured values of \Tone~for the MET. This result is also consistent with the limit of 
$4 \times 10^{-8}$ obtained by Kim \textit{et al.} through a similar
argument~\cite{PhysRevLett.106.120501}. Note that our small value for the loss tangent should be considered as an effective value in the single-photon limit and is valid only for frequencies unaffected by strongly coupled TLS resonances. The small volume of the tunnel junction is undoubtedly a key factor in avoiding problematic TLS interactions that are otherwise inevitable in bulk studies of the loss tangent.


\section{\label{sec:conclusion}Conclusion}

We have successfully demonstrated the operation of a merged element transmon in which the bulk of the shunt capacitance is due to the junction itself. While not yet optimized for charge noise due to their rather low values of \ejec, the devices still showed reasonably good \Tone\ and \Ttwo\ values. Three devices demonstrated mean $\Tone > 80 \ \us$, with instances of \Tone\ exceeding 100 \us\ for a period of hours. A simple scaling argument suggests that such good performance in a large-junction device implies that junction losses in small-junction devices are not the dominant limiting factor. Future work incorporating epitaxial dielectrics into a MET design could result in further improvement in performance, while also providing a possible pathway to greatly increase the areal density of superconducting qubits. 

\begin{acknowledgments}

We thank the staff at the IBM Microelectronics Research Lab and Central Scientific Services for device fabrication. We also thank Oliver Dial for helpful discussions.

\end{acknowledgments}

\providecommand{\noopsort}[1]{}\providecommand{\singleletter}[1]{#1}%

\end{document}